# Weak dissipation does not result in disappearance of persistent current

V.L.Gurtovoi, A.I. Ilin, A.V. Nikulov, and V.A. Tulin
Institute of Microelectronics Technology, Russian Academy of Sciences, 142432 Chernogolovka, Moscow District, Russia

Experimental results obtained last years corroborate a prediction made by I.O. Kulik forty years ago that the energy dissipation does not result in disappearance of equilibrium circular current observable in the normal state of superconductor rings and normal metal rings. Contrary interpretations of the persistent current as a Brownian motion or a dissipationless current are compared in the point of view of the observations of this phenomenon at presence of an electric potential difference. Distinctions between the quantum phenomena at atomic and mesoscopic levels are accentuated. In connection of the quantum oscillations in magnetic field of potential difference observed on asymmetric rings with the persistent current, it is pointed out that an experimental check of such phenomenon at thermodynamic equilibrium is possible.

**Introduction**
The experimental results obtained last years show that predictions of the persistent current circulating in a ring with nonzero resistance, made by I.O. Kulik forty years ago [1,2], may have fundamental importance. In the first work [1], obviously initiated by the well-known Aslamazov-Larkin theory [3] of fluctuation superconductivity, it has been shown that the persistent current can be observed not only in a superconducting state when the electric resistance is equal to zero, but also in the normal state when the resistance is not equal to zero. It was shown in the second work [2] that the persistent current state is possible without the superconducting long-rang order and consequently this quantum phenomenon can be observed in normal metal.

The possibility of the persistent current state is connected with the quantization $rp = n\hbar$ of the angular momentum $rp$, postulated by Bohr as far back as 1913 for the description of stability of electron orbits in atom. The permitted states of a free (not dissipating) electron being in an one-dimensional (with small section of the circle $s$) ring with radius $r$ should be discrete as well as in atomic orbits. Because of the relation $p = mv + qA$ between velocity $v$ and canonical momentum $p$ in the presence of a magnetic vector potential $A$, the permitted velocity

$$v_n = \frac{\hbar}{mr}\left(n - \frac{q\Phi}{2\pi\hbar}\right) = \frac{\hbar}{mr}\left(n - \frac{\Phi}{\Phi_0}\right) \qquad (1)$$

can not be equal to zero, when magnetic flux inside the ring $\Phi = 2\pi rA = B\pi r^2$ is not divisible by the flux quantum $\Phi_0 = 2\pi\hbar/q$. Electrons in a normal metal ring occupy the permitted levels $n = 0, 1, -1, 2, -2 \ldots n_F \approx mv_Fr/\hbar$, $-n_F \approx -mv_Fr/\hbar$ having opposite directed velocity (1). Therefore in an one-dimensional ring the value of the persistent current cannot exceed $I_p \approx ev_F/2\pi r$, where $v_F$ is the Fermi velocity [4]. This value is equal $I_p \approx 5 \; 10^{-8} \; A = 50 \; nA$ in a ring with diameter $2r \approx 1 \; \mu m$ and $v_F \approx 10^6 \; m/c$ typical for metals. In a real case the current value should be lower. It can be observed only at very low temperature $T$ when $k_BT$ does not exceed the energy difference between permitted levels $\Delta E_{n+1,n} = mv_{n+1}^2/2 - mv_n^2/2 = (2n+1)\hbar^2/2mr^2$ [2]. This difference $\Delta E_{1,0} = \hbar^2/2mr^2 \approx 2 \; 10^{-26} \; J$ between the bottom levels $n=1$ and $n=0$ of a ring in diameter $2r \approx 1 \; \mu m$ corresponds to very low temperature $\Delta E_{1,0}/k_B \approx 0.0016 \; K$ and is larger $\Delta E_{n_F+1,n_F} \approx \hbar^2 n_F/mr^2 \approx \hbar v_F/r \approx 2 \; 10^{-22} \; J$ on the Fermi level where $n_F$

$\approx mv_Fr/\hbar \gg 1$. Therefore the persistent current in metal rings can be observed at a temperature higher *0.0016 K*.

The persistent current in superconductor $I_p = s2en_sv_n = I_{p,A}(n - \Phi/\Phi_0)$ is much higher than in normal metal and it can be observed at any temperature $T \leq T_c$ because of the same quantum number $n$ of all $N_s = Vn_s = s2\pi rn_s$ pairs in the ring with minimal permitted energy $\propto v^2 \propto (n - \Phi/\Phi_0)^2$ [5]. A energy difference between permitted levels $\Delta E_{n+1,n}$ for superconducting condensate in $N_s$ times more than for electron because of the impossibility of any individual change of pair quantum number $n$ [6]. The number of pairs $N_s$ is huge in a real ring even in the fluctuations region at $T \geq T_c$ [3]. The persistent current was observed at $T \geq T_c$ of superconductor cylinder for some decades earlier, than in normal metal ring since the second task is much more difficult. It is more difficult to observe extremely weak current at extremely low temperature. The first experimental evidence of the velocity quantization (1) is the Little-Parks experiment made as far back as 1962 [7]. The first attempts to observe the persistent current in normal metal rings were made almost thirty years later [8] and only recently these efforts [9,10] have crowned success [11,12]. The authors [11,12] could observe the persistent current oscillations $I_p \approx I_{p,A}sin(2\pi\Phi/\Phi_0)$ with the amplitude $I_{p,A} = 0.001\ nA - 1\ nA$ and the period $\Phi_0 = 2\pi\hbar/e$ corresponding to the single electron charge $q = e$ in the temperature region $T = 0.03 - 3\ T_c$. Measurements [11] made on aluminium rings in radius $r \approx 308\ nm,\ 418\ nm$ and *793 nm* in magnetic fields $B > 5\ T$, much higher the critical field of superconducting transition of aluminium, have shown that the amplitude $I_{p,A}$ decreases exponentially with temperature $T$ and the ring radius $r$ increase. The dependencies $I_{p,A}(T,r)$ obtained at the measurements of the $I_p(\Phi/\Phi_0)$ oscillations [11] agree with the theoretical prediction for non-interacting diffusive electrons [13]. The measurements [14] made before have allowed to receive the temperature dependencies of the amplitude $I_{p,A}(T)$ of the persistent current oscillations $I_p(\Phi/\Phi_0)$ with the period $\Phi_0 = \pi\hbar/e$, corresponding to a superconducting pairs charge $q = 2e$ in the region of superconducting transition $T \approx T_c$. These measurements were made also on aluminium rings with $r \approx 350\ nm,\ 500\ nm,\ 1\ \mu m$ and $2\ \mu m$, but in low magnetic field $B < 0.01\ T$. The dependencies $I_{p,A}(T)$, obtained in [14], agree with the predictions of the fluctuations theory of one-dimensional superconductor [15].

**1. Why the persistent current does not decay?**
Thus, the possibility of the persistent current in the normal state [1] and normal metal [2], considered by I.O. Kulik forty years ago, has found full experimental corroboration [11,12,14]. The persistent current is observed in complete agreement with the theories basing on universally recognized principles of quantum mechanics. However, the nature of this very paradoxical quantum phenomenon remains mysterious. As the authors [11] note rightly, an electric current in a resistive circuit should rapidly decay in the absence of an electric field. For example, in the aluminium ring with $r \approx 308\ nm$, used in [11], the current should decay $I(t) = I_0exp(-t/\tau_{re})$ during the relaxation time $\tau_{re} = L/R \approx 1.5\ 10^{-12}\ c$, at its inductance $L \approx 3\ 10^{-12}\ H$ and resistance $R \approx 2\ Ohm$. But the persistent current can be observed permanently. There is no unequivocal answer to the question: "How can it be possible?" The authors [11] are sure that the persistent current flows through the resistive circuit without dissipating energy. But they do not try even to explain how a dissipationless current can be possible at finite value of electron mean free path. The exponential reduction of the oscillations $I_p \approx I_{p,A}sin(2\pi\Phi/\Phi_0)$ amplitude $I_{p,A}$ with the temperature increase, observed in [11], gives unequivocal evidence of an energy exchange between carriers of the persistent current and an environment. Therefore, as the author of the article [16] writes: "The idea that a normal, nonsuperconducting metal ring can sustain a persistent current - one that flows forever without dissipating energy - seems preposterous". However he, as well as the authors [11], interprets the persistent current as a dissipationless current. As opposed to the authors [11,16] I.O. Kulik, considering the persistent current in normal metal, emphasized as far back as 1970 [2]: "The current state corresponds in this case to a minimum of free energy, therefore the taking into account of a dissipation does not result in its disappearance". This prediction is corroborated with both the experiment [11] and theory [13]. The experimental dependencies $I_{p,A}(T, r)$ [11] agree with the theory [13] considering the persistent current in regime of diffusive transport, i.e. when the mean free path $l_e$ is less than the circle length $2\pi r$. The value $l_e \approx 4 \; 10^{-8}$ m = 40 nm is much less than the circle length $2\pi r \geq 2000$ nm at the diffusion constant $D = l_e v_F/3 \approx 0.025 \; m^2/c$, measured in [11], the Fermi velocity of aluminium $v_F \approx 2 \; 10^6$ m/c and the minimal ring radius $r \approx 308$ nm [11]. The experimental dependencies $I_{p,A}(T, r)$ [11] agree with the theoretical one [13] just at this value of the mean free path $l_e$. Because of such experimental data the confidence of the authors [11,16] in dissipationless nature of the persistent current looks entirely unfounded.

Any motion under equilibrium condition, i.e. when the free energy is minimal, at non-zero energy dissipation is well-known as Brownian motion [17]. Thus, I.O. Kulik stated [2] that the persistent current observed in a ring with nonzero resistance is a Brownian motion. The Brownian motion in a resistive electric circuit has been investigated as far back as 1928, experimentally by J. B. Johnson [18] and theoretically by H. Nyquist [19]. They have shown, that any resistance at a temperature $T$ "is noisy" with a power $W_{Ny} = 4k_BT\Delta f$ in any frequency band $\Delta f$ from zero up to the quantum limit $k_BT/2\pi\hbar$ [17]. This equilibrium phenomenon is known as Johnson's noise [17] or Nyquist noise [20]. The Nyquist noise does not decay at non-zero energy dissipation, as well as the persistent current, according to the I.O. Kulik's statement [2]. The Nyquist equilibrium current with the amplitude of the order $<I_{Ny}^2>^{1/2} \approx (4k_BT\Delta f/R)^{1/2}$ in a frequency band $\Delta f$ should be observed in a ring with a resistance $R$ along its circle. This amplitude $<I_{Ny}^2>^{1/2} \approx 200$ nA in the aluminium ring with $r \approx 308$ nm, $R \approx 2$ Ohm for whole frequency band $\Delta f \approx k_BT/2\pi\hbar$ at $T \approx 0.3$ K on two order more than the maximal persistent current observed in such ring [11]. But the persistent current as opposed to the Nyquist current is nonzero at zero frequency $\Delta f = f = 0$. This difference has fundamental importance, which, obviously, forces authors [11,16] to allege for the absence of any energy dissipation in the persistent current phenomenon.

It is obvious that the persistent current, as the directed equilibrium motion, can be observed only because of discreteness of a permitted state spectrum and due to the dependence of the momentum $p = mv + qA$ of a charged $q$ particle on the magnetic vector potential $A$. I.O. Kulik [2], as well as other authors [11,16], connect the second condition with the well-known Aharonov-Bohm effect. Y. Aharonov and D. Bohm in the famous work [21] published more than 50 years ago have paid attention to paradoxical effects connected with influence of electromagnetic potential on phase gradient $\nabla\varphi = p/\hbar = (mv + qA)/\hbar$ of wave function $\Psi=|\Psi|e^{i\varphi}$. The discreteness of the spectrum and the Aharonov-Bohm effect result in breach of symmetry between opposite directions under equilibrium condition. For example, at the magnetic flux inside a ring $\Phi = 0.25\Phi_0$ the state $n = 0$ with the velocity $v_0 = -0.25\hbar/mr$ is permitted (1) in the clockwise direction, for example, and it is forbidden in the anticlockwise one. The permitted state $n = 1$ with the velocity $v_0 = 0.75\hbar/mr$ (1) in the anticlockwise direction has a higher energy and, consequently, lower probability. The persistent current can be observed because of this probability difference of the motion in opposite direction. The breach of symmetry between opposite directions is observed in this phenomenon.

## 2. Stationary atomic orbits or Brownian motion in a system with discrete spectrum
Thus, different interpretations of the persistent current observed in rings with nonzero resistance are proposed. The authors [11] use the analogy to stationary electron orbits of atom as single argument for their interpretation of this phenomenon as a dissipationless current. Indeed, the persistent current, as well as stationary atomic orbits, can not be described without the Bohr's quantization $rp = n\hbar$.

Bohr in his famous work "On the constitution of atoms and molecules" [22] considered electron rotating around a nucleus with some velocity, which he has calculated. This velocity $v_{B1} = \hbar/m_e r_B \approx 2\ 10^6$ m/c on the first Bohr's orbit is close to the electron velocity $v_F \approx 2\ 10^6$ м/c, creating the persistent current in a normal metal ring. The analogy between the persistent current and stationary atomic orbits can seem almost full because of such concurrence. Many authors consider a loop with the persistent current as an artificial atom [23]. But as Werner Heisenberg noted rightly in his famous work [24], "*against formal rules, which are used in the quantum theory for calculation of observable parameters (for example, energy of hydrogen atom) serious objections are put forward*". The objections are connected with that "*these rules contain as an essential component the relation between parameters which, apparently, cannot be essentially observable (for example, position and time of electron rotation)*". Indeed, a notion about a real motion of electron in spherical-symmetric field of nucleus raises some questions, any answer on which results to contradiction with results of observations. For example, it is impossible to say about a trajectory and rotation direction of electron in spherical symmetric field. To avoid these insuperable difficulties Heisenberg has suggested to create "*bases of quantum mechanics which are founded on relations between essentially observable parameters*" [24]. Such approach has resulted in creation of the orthodox quantum mechanics, studied last eighty years. But some founders of the quantum theory, Planck, Einstein, Schrodinger, de Broglie and others have not accepted this change of the goal of scientific research. Instead of the description of real processes Heisenberg and Bohr have suggested to describe only results of observation. The debate of the founders of the quantum theory on the subject of its description has got a new urgency because of the famous works by John Bell's [25,26]. This philosophical debate became a subject of experimental researches [27] thanks to the Bell's theorem [25].

A question about a possibility of observations gets fundamental importance because the quantum mechanics can describe only observable parameters. Heisenberg, for example in the book [28], convincingly explaining why it is impossible to observe electron motion in atomic orbit, emphasized that there is no sense to speak about a direction of velocity and even the electron velocity by itself in this case. Not only this velocity, but also the angular momentum of atom cannot have a real direction. According to the well-known results of the Stern-Gerlach experiments the direction of angular momentum and of magnetic moment of atom can be considered only as a hidden variable. Bell proposed in the paper [26] a hidden variable model, describing realistically results of the Stern-Gerlach experiment. But in the work [25] he has shown that any realistic description, reproducing all predictions of observation results giving by the orthodox quantum mechanics, should assume a reality of a non-local interaction. In contrast to the atom case, there is no necessity at all to use a hidden variable for description of the persistent current in a ring. No experimental result forces us to doubt of the reality of observable parameters in this case. The ring, in contrast to atom, is not spherical-symmetric system. The current, circulating clockwise or anticlockwise along the ring, creates magnetic moment and angular momentum only in single direction, perpendicular to the ring plane. Just this real direction (clockwise or anticlockwise) of the persistent current is observed at measurements [11,12,14]. We should not doubt of the reality of this direction in contrast to the electron velocity on atomic orbit.

The authors [11,16] assert that "*time-reversal symmetry should forbid a current choosing one direction over the other around the ring*" and that "*the persistent current exists only in the presence of a magnetic field piercing the ring, which breaks time-reversal symmetry*". But such statement could make sense if only the direction of the persistent current $I_p$ changed with the direction of magnetic field. But as it is obvious from all experimental results [11,12,14] the $I_p$ direction is changes with he value of magnetic field. For example, if the persistent current is directed clockwise at the magnetic flux inside a ring $\Phi = 0.25\Phi_0$, then at $\Phi = 0.75\Phi_0$ it is directed anticlockwise. The observation of the $I_p$ direction change with the $\Phi$ value can reveal a more fundamental importance of the persistent current phenomenon, than it is assumed by the authors [11,16] ignoring this important experimental fact. In order to investigate a possibility of such change at atomic level, very high magnetic fields $\Phi_0/\pi r_B^2 \approx 4.7\ 10^9\ G$, inaccessible for the present, is needed. $r_B \approx 5.3\ 10^{-11}\ m$ is the radius of the first Bohr orbit. No effect observed at the atomic level up to now can be interpreted as an experimental evidence of symmetry breach between opposite directions. In contrast to the atomic level the breach of symmetry because of the Bohr's quantization and the Aharonov-Bohm effect is observed with evidence in the persistent current phenomenon.

In addition to this fundamental difference of the Bohr's quantization phenomena in atom and mesoscopic ring [29], there are also the others connected with difference of our experimental opportunities at these different levels of sizes. For example, we can create and measure a potential difference on the ring-halves with a nonzero resistance, passing through them an external current $I_{ext}$, as it is shown on Fig. 1. We can make also a ring with different section and, consequently, different resistance the ring-halves, Fig. 1. Such opportunities are important for experimental research of the nature of the persistent current. One of the obvious reasons of the confidence of the authors [11,16] in the dissipationless nature of the persistent current is the problem with the forces balance in the case of the opposite assumption. In the case of a conventional circular current $I$, when nobody doubts in the energy dissipation with a power $RI^2$, the average force, acting on electrons at their scattering, is compensated by the force $F_E = eE$ of electric field $E$. Therefore a current $I$, circulating in the ring with a resistance $R > 0$, can not decay for a long time at $RI = - d\Phi/dt = E2\pi r$. But the persistent current does not decay at magnetic flux $\Phi \neq n\Phi_0$ constant in time, i.e. without the Faraday's voltage $- d\Phi/dt = 0$. Therefore the assumption about energy dissipation in this case violates the force balance.

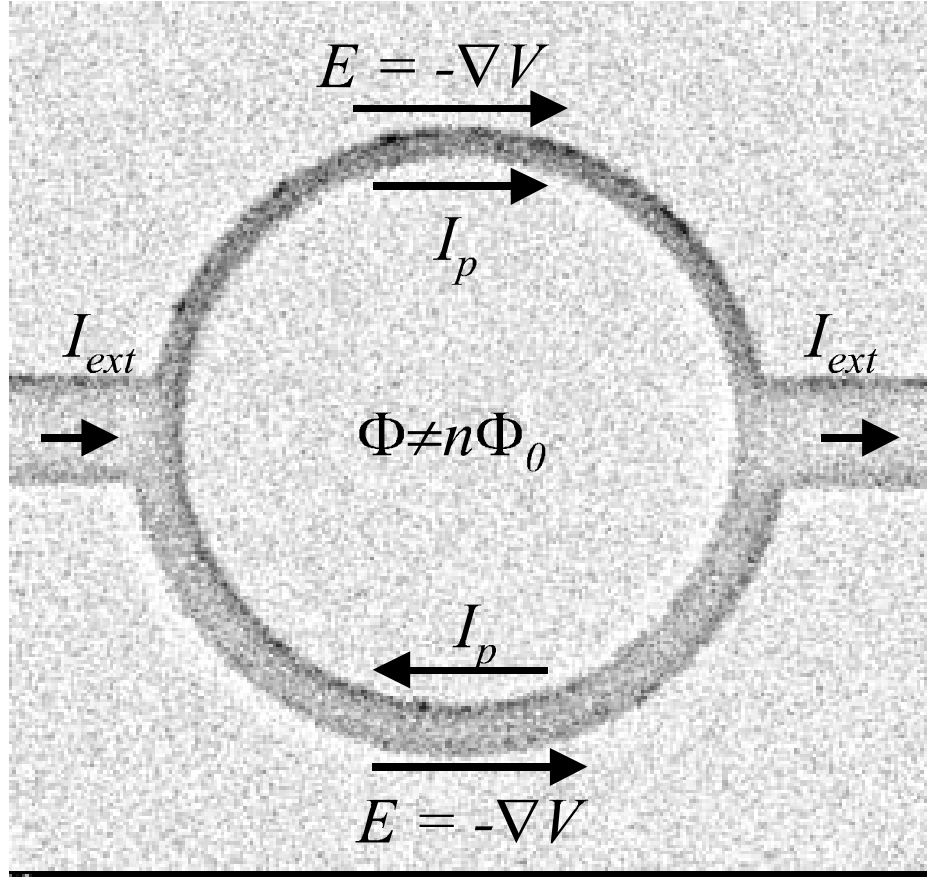

Fig. 1. An external current $I_{ext}$ creates a potential difference $V$ on the ring-halves with a nonzero resistance, passing through them. The observations of the Little-Parks oscillations $V(\Phi/\Phi_0) = R(\Phi/\Phi_0)I_{ext}$ prove that the persistent current does not decay at $V \neq 0$ and can flow against the direct electric field $E = -\nabla V$ in one of the ring-halves. An asymmetric ring with different section of the ring-halves is shown on which a sign-variable potential difference $V_p(\Phi/\Phi_0) \propto I_p(\Phi/\Phi_0)$ can be observed at $I_{ext} = 0$.

### 3. Observations of the persistent current at presence of an applied voltage.

The assumption [11,16] on the dissipationless nature of the persistent current cannot guarantee against the violation of the force balance if this quantum phenomenon can be observed at a potential difference on the ring-halves, Fig. 1. The Little-Parks oscillations of ring resistance $R(\Phi/\Phi_0)$ measured in the region of superconducting transition $T \approx T_c$ give experimental evidence of such challenge to the force balance. W. A. Little and R. D. Parks observed in [7] quantum periodicity in the resistance $\Delta R(\Phi/\Phi_0)$ of a superconducting cylinder at $T \approx T_c$, where $0 < R(T) < R_n$, and interpreted it as observation of quantum periodicity in its transition temperature $\Delta T_c(\Phi/\Phi_0)$, using the experimental relation $\Delta R(\Phi/\Phi_0) \approx [dR/d(T-T_c)]\Delta T_c(\Phi/\Phi_0)$. According to the universally recognized explanation [7,30] the $T_c$ value decreases $\Delta T_c(\Phi/\Phi_0) \propto -I_p^2(\Phi/\Phi_0)$ and the $R$ value increases $\Delta R(\Phi/\Phi_0) \propto I_p^2(\Phi/\Phi_0)$ when the superconducting state with zero velocity $v = 0$ is forbidden (1) and the persistent current $I_p$ is observed [14]. The ring resistance is found as the relation $R = V/I_{ext}$ of the voltage $V$ measured on the ring-halves to measuring direct current $I_{ext}$ passing along these ring-halves from left to right, Fig.1, or from right to left. The electric field $E = -\nabla V - dA/dt = -\nabla V$ is directed also from left to right, Fig.1, or from right to left in the both ring-halves because of the Faraday's voltage absence $dA/dt = 0$. Consequently, the persistent current $I_p$, circulating clockwise or anticlockwise, flows in one of the ring-halves against action of the force of the direct electric field $E = -\nabla V$, Fig. 1. At $I_{ext} < I_p$ the total direct current can flow against the direct electric field $E = -\nabla V$.

In order to observe the Little-Parks oscillations at $I_{ext} < I_p$ a system with great number of aluminium rings connected in series was used in the work [31]. All rings have the identical diameter $2r \approx 1.9 \ \mu m$ and the sections of the ring-halves $s_w = w_w d \approx 0.008 \ \mu m^2$ and $s_n = w_n d \approx 0.004 \ \mu m^2$ (film thickness $d = 20 \ nm$, width of the ring-halves $w_w \approx 0.4 \ \mu m$ and $w_n \approx 0.2 \ \mu m$). The amplitude of the $I_p(\Phi/\Phi_0)$ oscillations measured in the region of superconducting transition of similar aluminium rings [14] is equal $I_{p,A} \approx 100 \ nA$ at $T = T_c$. The Little-Parks oscillations $V(\Phi/\Phi_0) = R(\Phi/\Phi_0)I_{ext}$ could be observed on 110 rings at $I_{ext} \geq 50 \ нA$ in the work [31]. In addition to these oscillations of the resistance a sign-variable dc voltage $V_p(\Phi/\Phi_0)$ was observed at $I_{ext} = 0$ on the system of 110 asymmetric rings used in [31]. The sign of the dc voltage $V_p(\Phi/\Phi_0)$ observed at $I_{ext} = 0$ on the ring-halves with different sections $s_w \approx 0.008 \ \mu m^2 > s_n \approx 0.004 \ \mu m^2$ changes at $\Phi = n\Phi_0$ and $\Phi = (n+0.5)\Phi_0$ [31], as well as the direction of the persistent current $I_p(\Phi/\Phi_0)$ [11,12,14]. The appreciable oscillations $V_p(\Phi/\Phi_0)$ with amplitude $V_{p,A} \geq 50 \ nV$ were observed at temperatures $T \approx 1.34 \div 1.37 \ K$, corresponding to the bottom part of resistive transition $R \approx (0.03 \div 0.6)R_n$ [31]. The amplitude $V_{p,A}$ had a maximum value $\approx 600 \ nV$ at $T \approx 1.35 \ K$, corresponding to $R \approx 0.2R_n$ [31]. The Little-Parks oscillations $\Delta V(\Phi/\Phi_0) = \Delta R(\Phi/\Phi_0)I_{ext} \propto I_p^2(\Phi/\Phi_0)$, in contrast to the $V_p(\Phi/\Phi_0) \propto I_p(\Phi/\Phi_0)$ one, have minimum at $\Phi = n\Phi_0$ and maximum at $\Phi = (n+0.5)\Phi_0$. Therefore the oscillations $V(\Phi/\Phi_0) = R(\Phi/\Phi_0)I_{ext} + V_p(\Phi/\Phi_0)$ with these positions of the extremes, corresponding to the Little-Parks oscillations, are observed at $I_{ext} \geq 50 \ nA$ when the amplitude $\Delta R_A I_{ext} \geq 50 \ Ohm \ 50 \ nA = 2500 \ nV$ of the $\Delta R(\Phi/\Phi_0)I_{ext}$ oscillations exceeds noticeably the amplitude $V_{p,A} \leq 600 \ nV$ of the $V_p(\Phi/\Phi_0)$ oscillations [31].

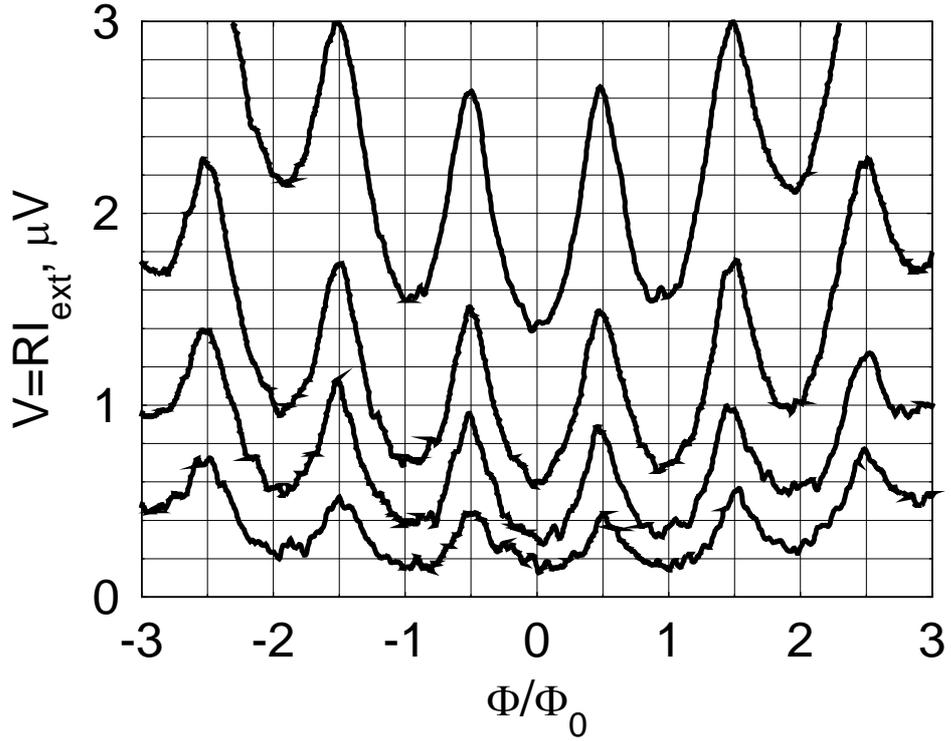

Fig. 2. The Little-Parks oscillations $V(\Phi/\Phi_0) = R(\Phi/\Phi_0)I_{ext}$ observed on the system of 1080 aluminium rings connected in series in diameter $2r \approx 2$ $\mu m$, the width of the ring-halves $w_w \approx 0.42$ $\mu m$ and $w_n \approx 0.26$ $\mu m$ and film thickness $d \approx 30$ $nm$, at different values of measuring current $I_{ext}$ and the temperature corresponding to the bottom part of the resistive transition. From below upwards: $I_{ext} = 0.2$ nA, $T \approx 1.3678$ K, $R \approx 0.075R_n$; $I_{ext} = 0.4$ nA, $T \approx 1.3684$ K, $R \approx 0.09R_n$; $I_{ext} = 0.6$ nA, $T \approx 1.3689$ K, $R \approx 0.12R_n$; $I_{ext} = 0.8$ nA, $T \approx 1.3710$ K, $R \approx 0.21R_n$. The potential difference measured at $\Phi = 0$ equals: $V \approx 120$ nV, (per one ring $V/1080 \approx 0.1$ nV); $V \approx 280$ nV, ($V/1080 \approx 0.26$ nV); $V \approx 600$ nV ($V/1080 \approx 0.55$ nV); $V \approx 1400$ nV ($V/1080 \approx 1.3$ nV). The resistance of the 1080 rings system of in the normal state $R_n \approx 8000$ Ohm, per one ring $R_{n1} = R_n/1080 \approx 7.4$ Ohm.

It is obvious, that the power $V_p^2/R$, observed at $I_{ext} = 0$ in the phenomenon of the quantum oscillations of the direct voltage $V_p(\Phi/\Phi_0) \propto I_p(\Phi/\Phi_0)$ [31], is induced by an uncontrollable noise which is in any measuring system. In order to observe the Little-Parks oscillations at measuring current $I_{ext} \geq 0.1$ nA, Fig. 2,3, much lower than it was possible to make in [31] we have diminished a level of the uncontrollable noise with help of industrial Pi-filters (Tusonix) and the distributed RC-systems used at cryogenic temperatures. We used also a system with number of rings 1080 ten times greater than in [31]. Measuring current $I_{ext} = 0.2 \div 0.8$ nA induces a potential difference $V = RI_{ext}$ on the ring-halves of the order nanovolt, Fig. 2. The observation of the higher resistance $\Delta R(\Phi/\Phi_0) \propto I_p^2(\Phi/\Phi_0)$ at $\Phi \neq n\Phi_0$, Fig. 2, proves that the persistent current $I_p(\Phi/\Phi_0)$ does not decay in spite of the electric field $E = -\nabla V$ directed against the electric current in one of the ring-halves, Fig.1. The amplitude of the $I_p(\Phi/\Phi_0)$ oscillations observed in similar aluminium rings [14] is much higher the measuring current $I_{p,A} \approx 100$ nA $>> I_{ext} = 0.2 \div 0.8$ nA. These experimental results, Fig. 2, prove that the unreasonable confidence of the authors [11,16] in a dissipationless nature of the persistent current cannot eliminate the challenge to the force balance. This confidence, regardless of measurement results, presupposes that the relaxation time $\tau_{re} = L/R$ of the persistent current is not very short $\approx 1.5$

$10^{-12}$ *c*, but infinity because of the zero resistance $R = 0$ of all rings used for the $I_p(\Phi/\Phi_0)$ oscillations observations [11,12,14]. But even this assumption about $R = 0$, contradicting to the experimental data, can not provide a reasonable description of the observation of the direct electric current $I_p$ flowing against the direct electric field $E = -\nabla V$. The resistance must be negative $R < 0$ in this case. An assumption about a negative resistance hardly can be accepted. Therefore it is necessary to search for other description of the phenomenon of the persistent current observed in rings with nonzero resistance.

**4. Switching between states with different connectivity of wave function.**
The description, proposed in [5] for the case of the Little-Parks experiment, proceeds from the experimental fact that the persistent current $I_p \neq 0$ at $R > 0$ is observed only in the critical region at $T \approx T_c$, where the resistant $0 < R(T) < R_n$ because of thermal fluctuations. Under equilibrium condition $I_p \neq 0$ but $R = 0$ at $T < T_c$, where fluctuations are neglecting small and $R = R_n$, but $I_p = 0$ at $T > T_c$, where fluctuations are also neglecting small. Thus, the Little-Parks oscillations $\Delta R(\Phi/\Phi_0) \propto I_p^2(\Phi/\Phi_0)$ should be considered as a fluctuation phenomenon. This experimental fact supported by the Kulik theory [1] gives an additional argument for interpretation of the persistent current $I_p \neq 0$ observed at $R > 0$ as a Brownian motion. The persistent current $I_p \neq 0$ and non-zero resistance $R > 0$ are incompatible under stationary condition. The discreteness of spectrum providing the possibility $I_p \neq 0$ can be only at the phase coherence along all circle, when the quantization condition $\oint_l dl \nabla \varphi = 2\pi n$ is valid. It is possible if only all segments of the ring are in the superconducting state or there is a Josephson connection between superconducting segments. The resistance equals to zero $R = 0$ in this case. Therefore $I_p \neq 0$ at $R > 0$ is observed only at $T \approx T_c$ where thermal fluctuations switch the ring between superconducting states with different connectivity of the wave function $\Psi = |\Psi|e^{i\varphi}$. When whole ring is in superconducting state, the angular momentum of each pair is $rp = rmv + 2e\Phi/2\pi = n\hbar$ because of the Bohr's quantuzation and the persistent current $I_p = s2en_sv_n = (s2en_s\hbar/rm)(n - \Phi/\Phi_0)$ circulates at $\Phi \neq n\Phi_0$, clockwise or anticlockwise. The condition of quantization $\oint_l dl \nabla \varphi = 2\pi n$ disappears at a transition of a ring segment in the normal state. The potential difference $V(t) = R_sI(t)$ should arise and the current should decay $I(t) = I_0exp(-t/\tau_{re})$ during the relaxation time $\tau_{re} = L/R_s$ because of a finite resistance $R_s > 0$ of the segment in the normal state. The velocity of each pair in the superconducting part of the ring decrease up to zero $v = 0$ and their angular momentum changes from $rp = n\hbar$ to $rp = 2e\Phi/2\pi$ under action of the force $F_E = 2eE(t)$ of the electric field $E(t) = \nabla V(t)$. This electric field should arise because of the energy dissipation with the power $V(t)I(t) = R_sI^2(t)$ in the segment switched in the normal state. Consequently the angular momentum of each pair changes from $rp = n\hbar$ to $rp = 2e\Phi/2\pi$ because of the dissipation force.

This change should be compensated because of the quantization $rp = n\hbar$ when the ring segment will return to the superconducting state. The ring should return in superconducting state with a quantum number $n$ when the wave function is closed in it and the quantization condition $\oint_l dl \nabla \varphi = 2\pi n$ becomes valid again. This quantum number should have with predominant probability the same integer value $n$ corresponding to the minimal energy because of the strong discreteness $\Delta E_{n+1,n} \gg k_BT$ of the permitted state spectrum. Just therefore the quantum oscillations of the average value of the persistent current $\langle I_p \rangle \propto (\langle n \rangle - \Phi/\Phi_0)$ are observed at $R > 0$ in the Little-Parks effect $\Delta R(\Phi/\Phi_0) \propto \langle I_p^2 \rangle \propto \langle (n - \Phi/\Phi_0)^2 \rangle$, Fig.2, and at magnetization measurements [14], $M \propto \langle I_p \rangle \propto (\langle n \rangle - \Phi/\Phi_0)$. Both $\langle I_p \rangle \propto (\langle n \rangle - \Phi/\Phi_0)$ and $\langle I_p^2 \rangle \propto \langle (n - \Phi/\Phi_0)^2 \rangle$ values equal zero at $\Phi = n\Phi_0$ when the single permitted state $n$ has minimum energy $\propto (n - \Phi/\Phi_0)^2 = 0$. But at $\Phi = (n+0.5)\Phi_0$ two states $n$ and $n+1$ with the opposite persistent current $\langle I_p \rangle \propto n - \Phi/\Phi_0 = -0.5$ and 0.5 have minimum energy $\propto (n - \Phi/\Phi_0)^2 = 0.25$. Therefore $<I_p> \propto (<n> - \Phi/\Phi_0) = 0$ [14] but $<I_p^2> \propto <(n - \Phi/\Phi_0)^2>$ has maximum value, Fig.2, at $\Phi = (n+0.5)\Phi_0$.

According to [5] the persistent current average in time $<I_p>$ can not decay in spite of the non-zero resistance average in time $<R> > 0$ because of the compensation of the deflection of angular momentum from the quantum value $rp = n\hbar$ under action of the dissipation force by its recurrence to this value $rp = n\hbar$ at the closing of the wave function in the ring. This recurrence in a time unit at its numerous repeating $N_{sw}$ during a long time $\Theta >> \tau_{re}$, at the switching of the ring between superconducting states with different connectivity of wave function was named in the article [5] quantum force. The angular momentum should change on $(2e\Phi/2\pi - <n>\hbar)N_{sw}/\Theta = \hbar(\Phi/\Phi_0 - <n>)\omega_{sw}$ in a time unit because of the dissipation at a switching frequency $\omega_{sw} = N_{sw}/\Theta << 1/\tau_{re}$ when the pair velocity has time to decrease down to zero between the switching. This change should be compensated by

$$rF_q = \hbar \left( <n> - \frac{\Phi}{\Phi_0} \right) \omega_{sw} \qquad (2)$$

because of the recurrence of the angular momentum to the quantum value $rp = n\hbar$. The quantum "force" $F_q$ is not potential, as well as Faraday's voltage $-d\Phi/dt$, and cannot be located in a ring segment. The angular "force" $rF_q$ (2), replacing $-2ed\Phi/dt$, restores the forces balance in the phenomenon of the persistent current $<I_p> \neq 0$ observed without any decay at $<R> > 0$.

J. E. Hirsch notes in [32] that "*an azimuthal quantum force acting on electrons only would change the total angular momentum of the system, violating the physical principle of angular momentum conservation*". This criticism hits far from the mark. The quantum "force" introduced in [5] does not explain, but only describes the phenomenon. The results of [5] can not apply for an explanation not only of the Meissner effect puzzle, but also of the Little-Parks effect. In the case of quantum effects it is not a shortcoming. There is important to remind that the orthodox quantum mechanics does not explain, but only describes quantum phenomena. For example, the Bohr quantization and the Schrodinger equation describe a discrete spectrum, but they cannot explain, why the spectrum is discrete. J.E. Hirsch in the papers [32,33] and others notes rightly on the puzzle of the Meissner effect. It is necessary to agree with his statement that the conventional theory of superconductivity cannot explain why charge carriers can accelerate against the Lorentz electric force at the Meissner effect [33]. And it is indeed very strange that "the question of what is the 'force' propelling the mobile charge carriers and the ions in the superconductor to move in direction opposite to the electromagnetic force in the Meissner effect was essentially never raised nor answered" [32], except for few instances. This puzzle, ignored by most physicists, is obvious not only in the case of the Meissner effect. The same puzzle is evident in the phenomenon of the persistent current [11,12,14,31] which does not decay at the taking into account of a dissipation [2]. Here it is necessary to emphasize, that I.O. Kulik [1,2] and the others [4,13,15] have described this phenomenon [1,2] on the basis of the orthodox quantum formalism and has connected it with the Aharonov-Bohm effect [21]. This effect right from the beginning [21] and till now [34,35] is a subject of discussions concerning not-local change of the phase gradient $\nabla\varphi$ [27], which is connected in quantum mechanics with the momentum of a quantum particle $\hbar\nabla\varphi = p$. The Aharonov-Bohm effect [21] implies a non-local force-free momentum transfer [27].

The orthodox quantum mechanics refuses to address such puzzle [27]. A realistic interpretation of the quantum theory in terms of hidden variables [36] suggested by Bohm as far back as 1952 contains a quantum potential which reveals that universally recognize quantum formalism presupposes non-local force-free momentum transfer at its realistic interpretation. The Bohm's theory [36] is well-known among experts in quantum mechanics foundation but poorly known outside this circle. David Mermin writes in the paper "Hidden variables and the two theorems of John Bell" [37] that *"Bell's favorite example of a hidden-variables theory, Bohm theory [36], is not only explicitly contextual but explicitly and spectacularly non-local"*. Bell in his famous work [25] has generalized the Bohm theory [36], proving that any realistic interpretation of the orthodox quantum mechanics presupposes a non-local interaction. J.E. Hirsch proposes in [32] a realistic description of the Meissner effect puzzle. But his explanation [32] based on the hole theory of superconductivity provokes some obvious objections. The puzzle of the force-free momentum transfer cannot be restricted to the Meissner effect or even superconductivity. The Aharonov-Bohm effect in the case of the two-slit interference experiment [27] can be described realistically [38] with the help of the Bohm quantum potential, which changes momentum of particles. But it is doubtful that the spectacularly non-local quantum potential may be acceptable as a real force. The non-local force-free momentum transfer implied in the Aharonov-Bohm effect ought be considered as an outstanding puzzle. We can for the present only use a description of this puzzle at considerations of quantum phenomena, as it is made in the case of the quantum force. It was shown in [6] that this puzzle is more real in the case of the Aharonov-Bohm effect in superconductors, than in the two-slit interference experiment. This difference may be connected with different essence [39] of the Ginzburg - Landau wave function describing the quite real density of superconducting pairs and the Schrodinger wave function in Born's interpretation, describing a probability density, which should collapse at observation [27].

**5. Could the potential difference $V_p(\Phi/\Phi_0)$ can be observed under thermodynamic equilibrium?**
I.O. Kulik's statement, that the taking into account of a dissipation does not result in the decay of the persistent current [2], means that this current is similar to the conventional circular current $I$ in a ring with $R > 0$, maintained by the Faraday's voltage $RI = - d\Phi/dt$. As it is well-known in the latter case a potential difference $V = 0.5(R_n - R_w)I$ should be observed on ring-halves with different resistance $R_n > R_w$. Its value cannot be large on a single ring at a small current, for example $I = 1\ nA$, equal to the maximal amplitude of the persistent current observed in [11]. Therefore in order to verify that the persistent current can create a potential difference, just as the conventional circular current creates it, a system with a great number connected in series should be used. For example, the conventional current $I = 1\ nA$, circulating in a single aluminium ring with $2r \approx 1\ \mu m$, should create $V = 0.5(R_n - R_w)I \approx 0.25\ nB$ on the ring-halves with different section $s_w = w_w d \approx 0.01\ \mu m^2$, $s_n = w_n d \approx 0.005\ \mu m^2$ and the resistance $R_n = \rho_{Al}\pi r/s_n \approx 1\ Ohm$, $R_w = \rho_{Al}\pi r/s_w \approx 0.5\ Ohm$. This voltage should increase to $V \approx 420\ nV$ at using a system with 1680 rings used in [11] and up to $V \approx 2\ mV$ at 10 million rings used at one of the first attempts [8] to observe the persistent current in normal metal [16]. Thus, at the modern development of nanotechnology there is a real opportunity to observe a potential difference connected with the persistent current, even if its value on some orders is less than in the case of the conventional current.

Such phenomenon may seem impossible because of some fundamental principle of physics. But the quantum oscillations of the dc voltage $V_p(\Phi/\Phi_0)$, similar to the $<I_p>(\Phi/\Phi_0)$ oscillations, observed on asymmetric aluminium rings near its superconducting transition [31], testify to a possibility of such phenomenon. The potential difference $V_p(\Phi/\Phi_0)$ can be observed on the ring-halves with different section $s_w > s_n$ [31] because of switching of ring segments between superconducting and normal state [40]. At $T < T_c$ such switching is possible only because of a nonequilibrium noise [41,42] or an external alternating current [43] with the amplitude exceeding the critical current at the temperature of measurement [44]. In the region of superconducting transition $T \approx T_c$ the switching occurs at thermodynamic equilibrium, without external influences, due to thermal fluctuations. Just therefore the Little-Parks oscillations [7] can be observed under condition close to the thermodynamic equilibrium, Fig. 2,3. In order to verify that the potential difference $V_p(\Phi/\Phi_0) \propto \langle I_p \rangle(\Phi/\Phi_0)$ can be induced not only by nonequilibrium noise but also thermal fluctuations this noise in the cryogenic part of the measuring system should be diminished and a structure with an enough great number of asymmetric rings connected in series should be used. The same structure consist of asymmetric rings can be used for the control of the noise level.

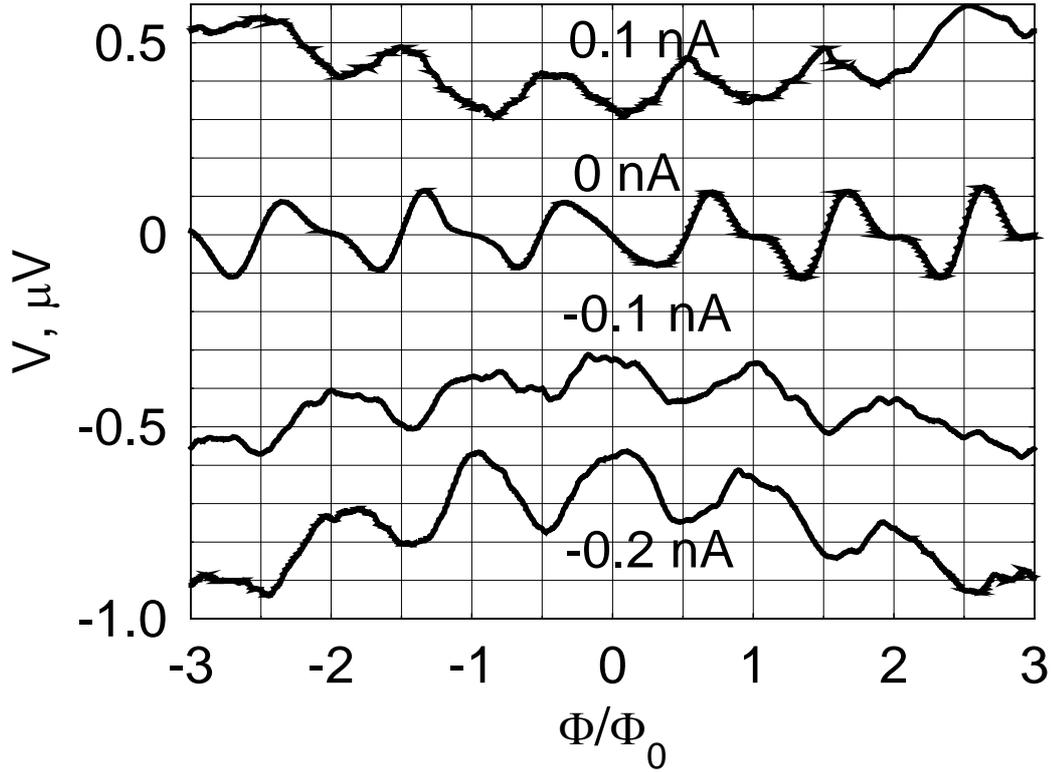

Fig. 3. The sign-variable oscillations of the direct voltage $V_p(\Phi/\Phi_0)$, induced by an uncontrollable noise diminished with help of industrial Pi-filters and the distributed RC-systems on system of 1080 rings at the temperature $T \approx 1.364$ K, corresponding to the bottom part of the resistive transition $R \approx 0.03R_n$ (0 nA) and the Little-Parks oscillations $V(\Phi/\Phi_0) = R(\Phi/\Phi_0)I_{ext}$ measured at a higher temperature $T \approx 1.374$ K, $R \approx 0.4R_n$ and the measuring current $I_{ext} = 0.1$ nA (0.1 nA), $I_{ext} = -0.1$ nA (-0.1 nA) and $I_{ext} = -0.2$ nA (-0.2 nA).

The observations of the $V_p(\Phi/\Phi_0)$ oscillations with the maximum amplitude $V_{A,max} \approx 15$ $\mu V$, on an individual loop in [41] and with $V_{A,max} \approx 0.6$ $\mu V$ on system of 110 rings connected in series in [31] testify, that the amplitude of uncontrollable noise $\langle I_{noise}^2 \rangle^{1/2}$ in the first case on some orders is more than in the second one. We could have measured the dependence of the amplitude $V_{A,max}$ of the $V_p(\Phi/\Phi_0)$ oscillations observed on the system of 110 rings on the amplitude $\langle I_{noise}^2 \rangle^{1/2}$ controllable noise down to $\langle I_{noise}^2 \rangle^{1/2} \approx 60$ nA thanks to the diminution of a level of uncontrollable noise in the cryogenic part of our measuring system with help of Pi-filters (Tusonix) and the distributed RC-systems. This calibration of the asymmetric rings system as a noise detector has allowed to estimate the amplitude $\langle I_{noise}^2 \rangle^{1/2} \approx 200$ nA of uncontrollable noise inducing on the system of 110 asymmetric aluminium rings the $V_p(\Phi/\Phi_0)$ oscillations with $V_{A,max} \approx 0.6$ $\mu V$ in the work [31]. In order to detect an uncontrollable noise diminished with help of Pi-filters and the distributed RC-systems we used a system of 1080 asymmetric aluminium rings. Our measurements have corroborate that the amplitude $V_{A,max}$ of the $V_p(\Phi/\Phi_0)$ oscillations induced by the controllable noise with the same amplitude $\langle I_{noise}^2 \rangle^{1/2}$ is approximately ten times greater on the system of 1080 rings than on the system of 110 similar rings. Our measurements have shown that an uncontrollable noise after the diminution of its level induces the $V_p(\Phi/\Phi_0)$ oscillations, Fig. 3, on the 1080 ring system with the maximum amplitude $V_{A,max} \approx 100\ nV$. This measurement result allows to conclude that Pi-filters (Tusonix) and the distributed RC-systems have diminished the amplitude of uncontrollable noise more than by the order, down to $<I_{noise}^2>^{1/2} \approx 10\ nA$. The power $W_{noise} = R_1(T)<I_{noise}^2> < R_{n,1}<I_{noise}^2> \approx 10^{-15}\ W$ of this noise per one ring with $R_1(T) < R_{n1} \approx 8\ Ohm$, corresponds to the power of the equilibrium noise $W_{Ny} = k_B T \Delta f$ at the temperature of measurement $T \approx 1.37\ K$ in the frequencies band $\Delta f \approx 50\ MHz$, which almost on three order smaller the quantum limit $k_B T/h \approx 30\ GHz$. The measurements made at a lower temperature [43] have shown that the $V_p(\Phi/\Phi_0)$ oscillations are induced irrespective of the frequency of an alternating current, at least, in the frequencies band 100 Hz - 1 MHz. These results show that we managed to come enough near to the equilibrium condition.

We managed to detect the $V_p(\Phi/\Phi_0)$ oscillations with period $\Phi_0/S$ corresponding to the ring area $S = \pi r^2 \approx 4\ \mu m^2$ down to its amplitude $V_A \geq 20\ nV$, using the Fourier transform of the measured dependencies $V_p(B)$. At such level of opportunities we could observe the $V_p(\Phi/\Phi_0)$ oscillations induced by the uncontrollable noise with the amplitude $<I_{noise}^2>^{1/2} \approx 10\ nA$ in the temperature region $T \approx 1.358 \div 1.372\ K$, corresponding to $R \approx (0.01 \div 0.25)R_n$. The Little-Parks oscillations are observed both at these temperatures, Fig.2, and at higher temperatures, Fig.3, corresponding to the top part of resistive transition, up to $R \approx R_n$. There is no valid reason to doubt that the $V_p(\Phi/\Phi_0)$ oscillations can be also observed at the higher temperatures. This observation can be made on a system with enough great number $N$ of rings. The proportionality $V_A \propto N$, corroborated by our measurements, gives an opportunity to observe a noticeable $V_p(\Phi/\Phi_0)$ oscillations with the amplitude $V_A \geq 20\ nV$ at any amplitude $V_A/N$ per one ring when the number $N$ of rings is enough great. The ring can give a maximum contribution to the voltage $V_p(\Phi/\Phi_0)$ at a temperature $T$ corresponding to the maximum of the $V_A(T-T_c)$ dependence. Therefore all rings of the system can give a contribution to $V_p(\Phi/\Phi_0)$ if only their critical temperature $T_c$ is the same. But the width of superconducting resistive transition $\approx 0.02\ K$ of the real aluminium systems, which we used, approximately in twenty times more than a width of the ideal transition determined only by thermal fluctuations. This means that the critical temperature of rings is scattered in an interval of temperatures $\approx 0.02\ K$, and only their twentieth part, i.e. $\approx 50$ from 1080, gives the contribution to $V_p(\Phi/\Phi_0)$, Fig.3, at the low level of noise $<I_{noise}^2>^{1/2} \approx 10\ nA$. Thus, $\approx 50$ asymmetric aluminim rings can detect the noise with the amplitude $<I_{noise}^2>^{1/2} \approx 10\ nA$. A like system of greater number of rings with more homogeneous $T_c$ can detect a weaker noise, down to the equilibrium one.

The results of our measurements and the made estimations testify to a real opportunity of observation of the potential difference $V_p(\Phi/\Phi_0) \propto <I_p>(\Phi/\Phi_0)$ under equilibrium conditions. The observation of this phenomenon will give a final confirmation of the interpretation of the persistent current observed at nonzero resistance as the direct Brownian motion. The confirmation of this interpretation, proposed by I.O. Kulik forty years ago [2], will have fundamental importance.

This work has been supported by a grant "Possible applications of new mesoscopic quantum effects for making of element basis of quantum computer, nanoelectronics and micro-system technics" of the Fundamental Research Program of ITCS department of RAS and the Russian Foundation of Basic Research grant 08-02-99042-r-ofi.